 \documentclass[preprint,prd,tightenlines, superscriptaddress]{revtex4-1}

\usepackage{graphicx} 
\usepackage{dcolumn}  
\usepackage{colordvi}
\usepackage{color}
\usepackage{epstopdf}
\usepackage{pstricks}
\usepackage{amssymb}
\usepackage{url}
\graphicspath{{ps}}
\usepackage{hyperref}
\usepackage{tabularx}
\usepackage{multirow}
\usepackage{units}
\usepackage{upgreek}
\usepackage{siunitx}
\usepackage{hyphenat}
\usepackage{subfigure}
\usepackage[italic]{hepnames}

\newcommand{\CP}{\ensuremath{C\hspace{-0.13em}P}\xspace}

\renewcommand{\Ph}{\HepParticle{h}{}{}\xspace}

\newcommand{\phitwo}{\ensuremath{\upphi_2}\xspace}

\renewcommand{\Prhozero}{\ensuremath{\HepParticle{\Prho}{}{}^0}\xspace}
\renewcommand{\PBzero}{\ensuremath{\HepParticle{\PB}{}{}^0}\xspace}
\renewcommand{\APBzero}{\ensuremath{\HepParticle{\APB}{}{}^0}\xspace}

\renewcommand{\APDzero}{\ensuremath{\HepParticle{\APD}{}{}^0}\xspace}
\renewcommand{\Pgpz}{\ensuremath{\HepParticle{\Pgp}{}{}^0}\xspace}
\renewcommand{\PDzero}{\ensuremath{\HepParticle{\PD}{}{}^0}\xspace}

\begin{document}

\def\belletwo {\it {Belle~II}}

\clubpenalty = 10000  
\widowpenalty = 10000 

\vspace*{-3\baselineskip}
\resizebox{!}{3cm}{\includegraphics{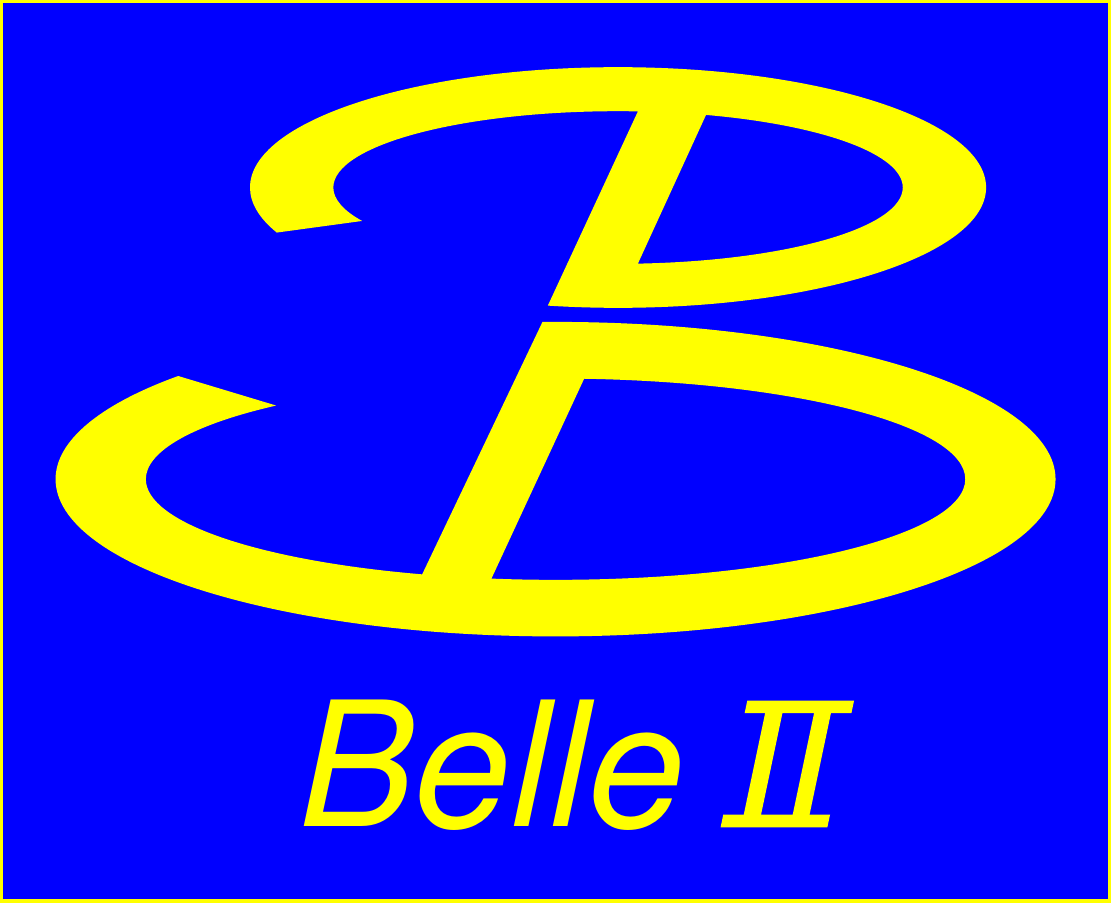}}

\vspace*{-5\baselineskip}
\begin{flushright}
BELLE2-CONF-PH-2021-003

\today
\end{flushright}

\quad\\[0.5cm]

\title {Angular analysis of $B^+ \to \rho^+\rho^0$ decays reconstructed in 2019--2020 Belle~II data}

\newcommand{\instCPPM}{Aix Marseille Universit\'{e}, CNRS/IN2P3, CPPM, 13288 Marseille, France}
\newcommand{\instBeihang}{Beihang University, Beijing 100191, China}
\newcommand{\instBNL}{Brookhaven National Laboratory, Upton, New York 11973, U.S.A.}
\newcommand{\instBINP}{Budker Institute of Nuclear Physics SB RAS, Novosibirsk 630090, Russian Federation}
\newcommand{\instCMU}{Carnegie Mellon University, Pittsburgh, Pennsylvania 15213, U.S.A.}
\newcommand{\instCinvestavIPN}{Centro de Investigacion y de Estudios Avanzados del Instituto Politecnico Nacional, Mexico City 07360, Mexico}
\newcommand{\instPrague}{Faculty of Mathematics and Physics, Charles University, 121 16 Prague, Czech Republic}
\newcommand{\instChiangMai}{Chiang Mai University, Chiang Mai 50202, Thailand}
\newcommand{\instChiba}{Chiba University, Chiba 263-8522, Japan}
\newcommand{\instChonnam}{Chonnam National University, Gwangju 61186, South Korea}
\newcommand{\instConacyt}{Consejo Nacional de Ciencia y Tecnolog\'{\i}a, Mexico City 03940, Mexico}
\newcommand{\instDESY}{Deutsches Elektronen--Synchrotron, 22607 Hamburg, Germany}
\newcommand{\instDuke}{Duke University, Durham, North Carolina 27708, U.S.A.}
\newcommand{\instITAR}{Institute of Theoretical and Applied Research (ITAR), Duy Tan University, Hanoi 100000, Vietnam}
\newcommand{\instRomaENEA}{ENEA Casaccia, I-00123 Roma, Italy}
\newcommand{\instEri}{Earthquake Research Institute, University of Tokyo, Tokyo 113-0032, Japan}
\newcommand{\instJuelich}{Forschungszentrum J\"{u}lich, 52425 J\"{u}lich, Germany}
\newcommand{\instFuJen}{Department of Physics, Fu Jen Catholic University, Taipei 24205, Taiwan}
\newcommand{\instFudan}{Key Laboratory of Nuclear Physics and Ion-beam Application (MOE) and Institute of Modern Physics, Fudan University, Shanghai 200443, China}
\newcommand{\instGoettingen}{II. Physikalisches Institut, Georg-August-Universit\"{a}t G\"{o}ttingen, 37073 G\"{o}ttingen, Germany}
\newcommand{\instGifu}{Gifu University, Gifu 501-1193, Japan}
\newcommand{\instSOKENDAI}{The Graduate University for Advanced Studies (SOKENDAI), Hayama 240-0193, Japan}
\newcommand{\instGyeongsang}{Gyeongsang National University, Jinju 52828, South Korea}
\newcommand{\instHanyang}{Department of Physics and Institute of Natural Sciences, Hanyang University, Seoul 04763, South Korea}
\newcommand{\instKEK}{High Energy Accelerator Research Organization (KEK), Tsukuba 305-0801, Japan}
\newcommand{\instJPARC}{J-PARC Branch, KEK Theory Center, High Energy Accelerator Research Organization (KEK), Tsukuba 305-0801, Japan}
\newcommand{\instHSE}{National Research University Higher School of Economics, Moscow 101000, Russian Federation}
\newcommand{\instIISER}{Indian Institute of Science Education and Research Mohali, SAS Nagar, 140306, India}
\newcommand{\instIITBhubaneswar}{Indian Institute of Technology Bhubaneswar, Satya Nagar 751007, India}
\newcommand{\instIITGuwahati}{Indian Institute of Technology Guwahati, Assam 781039, India}
\newcommand{\instIITHyderabad}{Indian Institute of Technology Hyderabad, Telangana 502285, India}
\newcommand{\instIITMadras}{Indian Institute of Technology Madras, Chennai 600036, India}
\newcommand{\instIndiana}{Indiana University, Bloomington, Indiana 47408, U.S.A.}
\newcommand{\instIHEPRussia}{Institute for High Energy Physics, Protvino 142281, Russian Federation}
\newcommand{\instHEPHYVienna}{Institute of High Energy Physics, Vienna 1050, Austria}
\newcommand{\instHiroshima}{Hiroshima University, Higashi-Hiroshima, Hiroshima 739-8530, Japan}
\newcommand{\instIHEPChina}{Institute of High Energy Physics, Chinese Academy of Sciences, Beijing 100049, China}
\newcommand{\instIPP}{Institute of Particle Physics (Canada), Victoria, British Columbia V8W 2Y2, Canada}
\newcommand{\instIOP}{Institute of Physics, Vietnam Academy of Science and Technology (VAST), Hanoi, Vietnam}
\newcommand{\instIFIC}{Instituto de Fisica Corpuscular, Paterna 46980, Spain}
\newcommand{\instFrascati}{INFN Laboratori Nazionali di Frascati, I-00044 Frascati, Italy}
\newcommand{\instNapoliINFN}{INFN Sezione di Napoli, I-80126 Napoli, Italy}
\newcommand{\instPadovaINFN}{INFN Sezione di Padova, I-35131 Padova, Italy}
\newcommand{\instPerugiaINFN}{INFN Sezione di Perugia, I-06123 Perugia, Italy}
\newcommand{\instPisaINFN}{INFN Sezione di Pisa, I-56127 Pisa, Italy}
\newcommand{\instRomaINFN}{INFN Sezione di Roma, I-00185 Roma, Italy}
\newcommand{\instRomaTreINFN}{INFN Sezione di Roma Tre, I-00146 Roma, Italy}
\newcommand{\instTorinoINFN}{INFN Sezione di Torino, I-10125 Torino, Italy}
\newcommand{\instTriesteINFN}{INFN Sezione di Trieste, I-34127 Trieste, Italy}
\newcommand{\instJAEA}{Advanced Science Research Center, Japan Atomic Energy Agency, Naka 319-1195, Japan}
\newcommand{\instMainz}{Johannes Gutenberg-Universit\"{a}t Mainz, Institut f\"{u}r Kernphysik, D-55099 Mainz, Germany}
\newcommand{\instGiessen}{Justus-Liebig-Universit\"{a}t Gie\ss{}en, 35392 Gie\ss{}en, Germany}
\newcommand{\instKarlsruhe}{Institut f\"{u}r Experimentelle Teilchenphysik, Karlsruher Institut f\"{u}r Technologie, 76131 Karlsruhe, Germany}
\newcommand{\instISU}{Iowa State University, Ames, Iowa 50011, U.S.A.}
\newcommand{\instKitasato}{Kitasato University, Sagamihara 252-0373, Japan}
\newcommand{\instKISTI}{Korea Institute of Science and Technology Information, Daejeon 34141, South Korea}
\newcommand{\instKoreaUnivKU}{Korea University, Seoul 02841, South Korea}
\newcommand{\instKSU}{Kyoto Sangyo University, Kyoto 603-8555, Japan}
\newcommand{\instKyungpook}{Kyungpook National University, Daegu 41566, South Korea}
\newcommand{\instLPI}{P.N. Lebedev Physical Institute of the Russian Academy of Sciences, Moscow 119991, Russian Federation}
\newcommand{\instLNNU}{Liaoning Normal University, Dalian 116029, China}
\newcommand{\instLMU}{Ludwig Maximilians University, 80539 Munich, Germany}
\newcommand{\instLuther}{Luther College, Decorah, Iowa 52101, U.S.A.}
\newcommand{\instMNITJaipur}{Malaviya National Institute of Technology Jaipur, Jaipur 302017, India}
\newcommand{\instMPP}{Max-Planck-Institut f\"{u}r Physik, 80805 M\"{u}nchen, Germany}
\newcommand{\instMPGHLL}{Semiconductor Laboratory of the Max Planck Society, 81739 M\"{u}nchen, Germany}
\newcommand{\instMcGill}{McGill University, Montr\'{e}al, Qu\'{e}bec, H3A 2T8, Canada}
\newcommand{\instMEPhI}{Moscow Physical Engineering Institute, Moscow 115409, Russian Federation}
\newcommand{\instNagoya}{Graduate School of Science, Nagoya University, Nagoya 464-8602, Japan}
\newcommand{\instNagoyaIAR}{Institute for Advanced Research, Nagoya University, Nagoya 464-8602, Japan}
\newcommand{\instNagoyaKMI}{Kobayashi-Maskawa Institute, Nagoya University, Nagoya 464-8602, Japan}
\newcommand{\instNaraWu}{Nara Women's University, Nara 630-8506, Japan}
\newcommand{\instNTUTaiwan}{Department of Physics, National Taiwan University, Taipei 10617, Taiwan}
\newcommand{\instNUUTaiwan}{National United University, Miao Li 36003, Taiwan}
\newcommand{\instKrakow}{H. Niewodniczanski Institute of Nuclear Physics, Krakow 31-342, Poland}
\newcommand{\instNiigata}{Niigata University, Niigata 950-2181, Japan}
\newcommand{\instNSU}{Novosibirsk State University, Novosibirsk 630090, Russian Federation}
\newcommand{\instOkinawa}{Okinawa Institute of Science and Technology, Okinawa 904-0495, Japan}
\newcommand{\instOsakaCity}{Osaka City University, Osaka 558-8585, Japan}
\newcommand{\instRCNP}{Research Center for Nuclear Physics, Osaka University, Osaka 567-0047, Japan}
\newcommand{\instPNNL}{Pacific Northwest National Laboratory, Richland, Washington 99352, U.S.A.}
\newcommand{\instPanjab}{Panjab University, Chandigarh 160014, India}
\newcommand{\instPanjabPAU}{Punjab Agricultural University, Ludhiana 141004, India}
\newcommand{\instRIKENMSL}{Meson Science Laboratory, Cluster for Pioneering Research, RIKEN, Saitama 351-0198, Japan}
\newcommand{\instSeoul}{Seoul National University, Seoul 08826, South Korea}
\newcommand{\instSPU}{Showa Pharmaceutical University, Tokyo 194-8543, Japan}
\newcommand{\instSoochow}{Soochow University, Suzhou 215006, China}
\newcommand{\instSoongsil}{Soongsil University, Seoul 06978, South Korea}
\newcommand{\instLjubljanaJSI}{J. Stefan Institute, 1000 Ljubljana, Slovenia}
\newcommand{\instKyiv}{Taras Shevchenko National Univ. of Kiev, Kiev, Ukraine}
\newcommand{\instTata}{Tata Institute of Fundamental Research, Mumbai 400005, India}
\newcommand{\instTUM}{Department of Physics, Technische Universit\"{a}t M\"{u}nchen, 85748 Garching, Germany}
\newcommand{\instTelAviv}{Tel Aviv University, School of Physics and Astronomy, Tel Aviv, 69978, Israel}
\newcommand{\instToho}{Toho University, Funabashi 274-8510, Japan}
\newcommand{\instTohoku}{Department of Physics, Tohoku University, Sendai 980-8578, Japan}
\newcommand{\instTitech}{Tokyo Institute of Technology, Tokyo 152-8550, Japan}
\newcommand{\instTokyoMetropolitan}{Tokyo Metropolitan University, Tokyo 192-0397, Japan}
\newcommand{\instUAS}{Universidad Autonoma de Sinaloa, Sinaloa 80000, Mexico}
\newcommand{\instNapoliUNIV}{Dipartimento di Scienze Fisiche, Universit\`{a} di Napoli Federico II, I-80126 Napoli, Italy}
\newcommand{\instPadovaUNIV}{Dipartimento di Fisica e Astronomia, Universit\`{a} di Padova, I-35131 Padova, Italy}
\newcommand{\instPerugiaUNIV}{Dipartimento di Fisica, Universit\`{a} di Perugia, I-06123 Perugia, Italy}
\newcommand{\instPisaUNIV}{Dipartimento di Fisica, Universit\`{a} di Pisa, I-56127 Pisa, Italy}
\newcommand{\instRomaTreUNIV}{Dipartimento di Matematica e Fisica, Universit\`{a} di Roma Tre, I-00146 Roma, Italy}
\newcommand{\instTorinoUNIV}{Dipartimento di Fisica, Universit\`{a} di Torino, I-10125 Torino, Italy}
\newcommand{\instTriesteUNIV}{Dipartimento di Fisica, Universit\`{a} di Trieste, I-34127 Trieste, Italy}
\newcommand{\instMontreal}{Universit\'{e} de Montr\'{e}al, Physique des Particules, Montr\'{e}al, Qu\'{e}bec, H3C 3J7, Canada}
\newcommand{\instIJCLab}{Universit\'{e} Paris-Saclay, CNRS/IN2P3, IJCLab, 91405 Orsay, France}
\newcommand{\instIPHC}{Universit\'{e} de Strasbourg, CNRS, IPHC, UMR 7178, 67037 Strasbourg, France}
\newcommand{\instAdelaide}{Department of Physics, University of Adelaide, Adelaide, South Australia 5005, Australia}
\newcommand{\instBonn}{University of Bonn, 53115 Bonn, Germany}
\newcommand{\instUBC}{University of British Columbia, Vancouver, British Columbia, V6T 1Z1, Canada}
\newcommand{\instCincinnati}{University of Cincinnati, Cincinnati, Ohio 45221, U.S.A.}
\newcommand{\instFlorida}{University of Florida, Gainesville, Florida 32611, U.S.A.}
\newcommand{\instHawaii}{University of Hawaii, Honolulu, Hawaii 96822, U.S.A.}
\newcommand{\instHeidelberg}{University of Heidelberg, 68131 Mannheim, Germany}
\newcommand{\instLjubljanaUniLJ}{Faculty of Mathematics and Physics, University of Ljubljana, 1000 Ljubljana, Slovenia}
\newcommand{\instLouisville}{University of Louisville, Louisville, Kentucky 40292, U.S.A.}
\newcommand{\instMalaya}{National Centre for Particle Physics, University Malaya, 50603 Kuala Lumpur, Malaysia}
\newcommand{\instLjubljanaUM}{Faculty of Chemistry and Chemical Engineering, University of Maribor, 2000 Maribor, Slovenia}
\newcommand{\instMelbourne}{School of Physics, University of Melbourne, Victoria 3010, Australia}
\newcommand{\instMississippi}{University of Mississippi, University, Mississippi 38677, U.S.A.}
\newcommand{\instUOM}{University of Miyazaki, Miyazaki 889-2192, Japan}
\newcommand{\instPittsburgh}{University of Pittsburgh, Pittsburgh, Pennsylvania 15260, U.S.A.}
\newcommand{\instUSTC}{University of Science and Technology of China, Hefei 230026, China}
\newcommand{\instSAlabama}{University of South Alabama, Mobile, Alabama 36688, U.S.A.}
\newcommand{\instSCarolina}{University of South Carolina, Columbia, South Carolina 29208, U.S.A.}
\newcommand{\instSydney}{School of Physics, University of Sydney, New South Wales 2006, Australia}
\newcommand{\instUTokyo}{Department of Physics, University of Tokyo, Tokyo 113-0033, Japan}
\newcommand{\instIPMU}{Kavli Institute for the Physics and Mathematics of the Universe (WPI), University of Tokyo, Kashiwa 277-8583, Japan}
\newcommand{\instVictoria}{University of Victoria, Victoria, British Columbia, V8W 3P6, Canada}
\newcommand{\instVPI}{Virginia Polytechnic Institute and State University, Blacksburg, Virginia 24061, U.S.A.}
\newcommand{\instWayneState}{Wayne State University, Detroit, Michigan 48202, U.S.A.}
\newcommand{\instYamagata}{Yamagata University, Yamagata 990-8560, Japan}
\newcommand{\instYerevan}{Alikhanyan National Science Laboratory, Yerevan 0036, Armenia}
\newcommand{\instYonsei}{Yonsei University, Seoul 03722, South Korea}
\newcommand{\instZZU}{Zhengzhou University, Zhengzhou 450001, China}
\affiliation{\instCPPM}
\affiliation{\instBeihang}
\affiliation{\instBNL}
\affiliation{\instBINP}
\affiliation{\instCMU}
\affiliation{\instCinvestavIPN}
\affiliation{\instPrague}
\affiliation{\instChiangMai}
\affiliation{\instChiba}
\affiliation{\instChonnam}
\affiliation{\instConacyt}
\affiliation{\instDESY}
\affiliation{\instDuke}
\affiliation{\instITAR}
\affiliation{\instRomaENEA}
\affiliation{\instEri}
\affiliation{\instJuelich}
\affiliation{\instFuJen}
\affiliation{\instFudan}
\affiliation{\instGoettingen}
\affiliation{\instGifu}
\affiliation{\instSOKENDAI}
\affiliation{\instGyeongsang}
\affiliation{\instHanyang}
\affiliation{\instKEK}
\affiliation{\instJPARC}
\affiliation{\instHSE}
\affiliation{\instIISER}
\affiliation{\instIITBhubaneswar}
\affiliation{\instIITGuwahati}
\affiliation{\instIITHyderabad}
\affiliation{\instIITMadras}
\affiliation{\instIndiana}
\affiliation{\instIHEPRussia}
\affiliation{\instHEPHYVienna}
\affiliation{\instHiroshima}
\affiliation{\instIHEPChina}
\affiliation{\instIPP}
\affiliation{\instIOP}
\affiliation{\instIFIC}
\affiliation{\instFrascati}
\affiliation{\instNapoliINFN}
\affiliation{\instPadovaINFN}
\affiliation{\instPerugiaINFN}
\affiliation{\instPisaINFN}
\affiliation{\instRomaINFN}
\affiliation{\instRomaTreINFN}
\affiliation{\instTorinoINFN}
\affiliation{\instTriesteINFN}
\affiliation{\instJAEA}
\affiliation{\instMainz}
\affiliation{\instGiessen}
\affiliation{\instKarlsruhe}
\affiliation{\instISU}
\affiliation{\instKitasato}
\affiliation{\instKISTI}
\affiliation{\instKoreaUnivKU}
\affiliation{\instKSU}
\affiliation{\instKyungpook}
\affiliation{\instLPI}
\affiliation{\instLNNU}
\affiliation{\instLMU}
\affiliation{\instLuther}
\affiliation{\instMNITJaipur}
\affiliation{\instMPP}
\affiliation{\instMPGHLL}
\affiliation{\instMcGill}
\affiliation{\instMEPhI}
\affiliation{\instNagoya}
\affiliation{\instNagoyaIAR}
\affiliation{\instNagoyaKMI}
\affiliation{\instNaraWu}
\affiliation{\instNTUTaiwan}
\affiliation{\instNUUTaiwan}
\affiliation{\instKrakow}
\affiliation{\instNiigata}
\affiliation{\instNSU}
\affiliation{\instOkinawa}
\affiliation{\instOsakaCity}
\affiliation{\instRCNP}
\affiliation{\instPNNL}
\affiliation{\instPanjab}
\affiliation{\instPanjabPAU}
\affiliation{\instRIKENMSL}
\affiliation{\instSeoul}
\affiliation{\instSPU}
\affiliation{\instSoochow}
\affiliation{\instSoongsil}
\affiliation{\instLjubljanaJSI}
\affiliation{\instKyiv}
\affiliation{\instTata}
\affiliation{\instTUM}
\affiliation{\instTelAviv}
\affiliation{\instToho}
\affiliation{\instTohoku}
\affiliation{\instTitech}
\affiliation{\instTokyoMetropolitan}
\affiliation{\instUAS}
\affiliation{\instNapoliUNIV}
\affiliation{\instPadovaUNIV}
\affiliation{\instPerugiaUNIV}
\affiliation{\instPisaUNIV}
\affiliation{\instRomaTreUNIV}
\affiliation{\instTorinoUNIV}
\affiliation{\instTriesteUNIV}
\affiliation{\instMontreal}
\affiliation{\instIJCLab}
\affiliation{\instIPHC}
\affiliation{\instAdelaide}
\affiliation{\instBonn}
\affiliation{\instUBC}
\affiliation{\instCincinnati}
\affiliation{\instFlorida}
\affiliation{\instHawaii}
\affiliation{\instHeidelberg}
\affiliation{\instLjubljanaUniLJ}
\affiliation{\instLouisville}
\affiliation{\instMalaya}
\affiliation{\instLjubljanaUM}
\affiliation{\instMelbourne}
\affiliation{\instMississippi}
\affiliation{\instUOM}
\affiliation{\instPittsburgh}
\affiliation{\instUSTC}
\affiliation{\instSAlabama}
\affiliation{\instSCarolina}
\affiliation{\instSydney}
\affiliation{\instUTokyo}
\affiliation{\instIPMU}
\affiliation{\instVictoria}
\affiliation{\instVPI}
\affiliation{\instWayneState}
\affiliation{\instYamagata}
\affiliation{\instYerevan}
\affiliation{\instYonsei}
\affiliation{\instZZU}
  \author{F.~Abudin{\'e}n}\affiliation{\instTriesteINFN} 
  \author{I.~Adachi}\affiliation{\instKEK}\affiliation{\instSOKENDAI} 
  \author{R.~Adak}\affiliation{\instFudan} 
  \author{K.~Adamczyk}\affiliation{\instKrakow} 
  \author{P.~Ahlburg}\affiliation{\instBonn} 
  \author{J.~K.~Ahn}\affiliation{\instKoreaUnivKU} 
  \author{H.~Aihara}\affiliation{\instUTokyo} 
  \author{N.~Akopov}\affiliation{\instYerevan} 
  \author{A.~Aloisio}\affiliation{\instNapoliUNIV}\affiliation{\instNapoliINFN} 
  \author{F.~Ameli}\affiliation{\instRomaINFN} 
% ...


\begin{abstract}
We report on the first Belle II measurement of the branching fraction~($\mathcal{B}$) and longitudinal polarization fraction ~($f_L$) of $B^+\to \rho^+\rho^0$ decays.  We reconstruct $B^+\to \rho^+(\to \pi^+\pi^0(\to \gamma\gamma))\rho^0(\to \pi^+\pi^-)$ decays in a sample of SuperKEKB electron-positron collisions collected by the Belle II experiment in 2019 and 2020 at the $\Upsilon$(4S) resonance and corresponding to $62.8$\,fb$^{-1}$ of integrated luminosity.  We fit the distributions of the difference between expected and observed $B$ candidate energy, continuum-suppression variable, dipion masses, and angular distributions of the resulting samples, to determine a signal yield of $104\pm16$ events. The signal yields are corrected for efficiencies determined from simulation and control data samples to obtain 
\begin{center}
$\mathcal{B}(B^+ \to \rho^+\rho^0) = [20.6 \pm 3.2(\rm stat) \pm 4.0(\rm syst)]\times 10^{-6}$, \mbox{and}  \\
\end{center}
\begin{center}
$f_L(B^+ \to \rho^+\rho^0) = 0.936  ^{+0.049}_{-0.041}(\rm stat)\pm 0.021(\rm syst)$.
\end{center}

This first Belle II $B^+ \to \rho^+\rho^0$ angular analysis yields results compatible with previous determinations, and indicates Belle II performance superior to early Belle results.
 
\end{abstract}


\maketitle

{\renewcommand{\thefootnote}{\fnsymbol{footnote}}}
\setcounter{footnote}{0}



\section{Introduction and motivation}

The study of charmless $B$ decays is a keystone of the worldwide flavor program. Processes mediated by $b\to u\bar{u}d$ transitions offer direct access to the unitarity angle \mbox{$\phitwo/\alpha \equiv \arg \left[\frac{-V_{td}V^*_{tb}}{V_{ud}V^*_{ub}}\right]$}, where $V_{ij}$ are elements of the quark-mixing matrix. In addition, they probe contributions of non-standard-model dynamics in processes mediated by amplitudes involving internal exchanges of $W$ bosons or heavy quarks (loops)~\cite{Kou:2018nap}.  However, a reliable extraction of weak phases and an unambiguous interpretation of measurements involving loop amplitudes in non-leptonic $B$ decays are spoiled by hadronic uncertainties,  which are rarely tractable in perturbative  calculations. Appropriately chosen combinations of measurements from decay modes related by flavor symmetries are used to reduce the impact of such  unknowns. An especially  fruitful approach consists in combining measurements of decays related by isospin symmetries. In particular, the combined analysis of branching fractions and charge-parity (\CP) violating asymmetries of the complete set of   $B^{+,0} \to \rho^{+,0}\rho^{0,-}$ isospin partners  enables a determination of~$\alpha$~\cite{Gronau:1990ka}. Belle~II, which has the  {\it unique}  capability  of  studying  jointly,  and within a consistent experimental environment, all relevant final states, is particularly promising for future determinations of~$\alpha$.

\begin{figure}[h!]
 \centering
    \centering
    \includegraphics[width=0.5\textwidth]{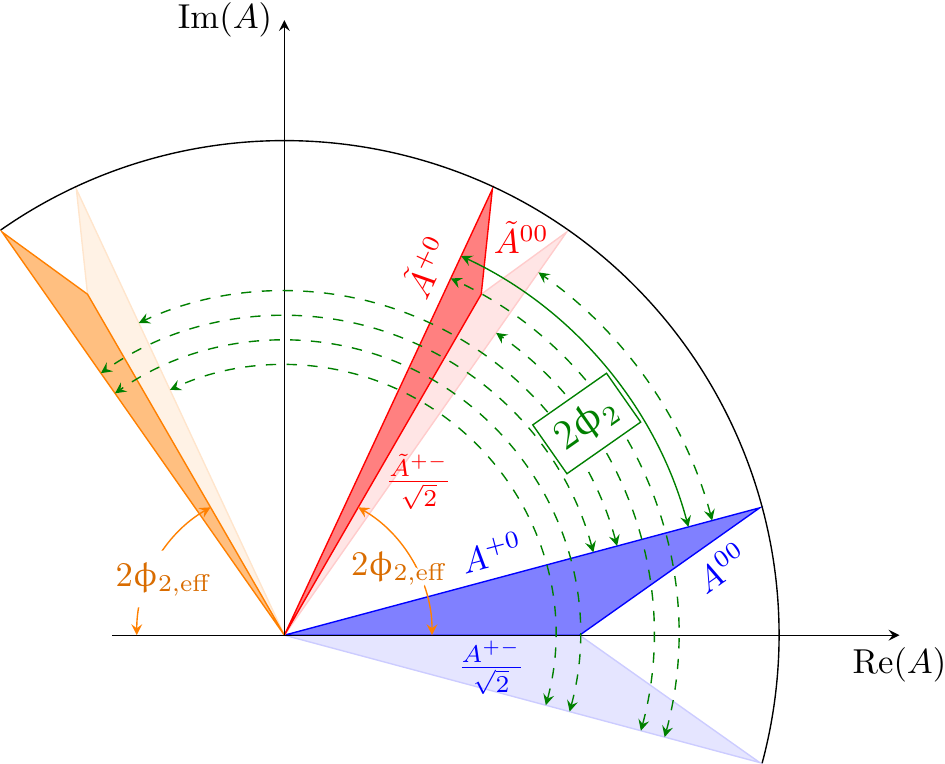}
 \caption{Geometrical representation of the isospin triangular relations in the complex plane of $\PB^{i+j}\to \Ph^i\Ph^j$ amplitudes. The blue and red shaded regions correspond to the isospin triangles. The angle between the \CP conjugate charged amplitudes   $A^{+-}$ and $\tilde{A}^{+-}$ corresponds to twice the weak phase $\alpha_{\rm eff}$ (orange solid lines). The angle between the \CP conjugate charged amplitudes $A^{+0}$ and $\tilde{A}^{+0}$ corresponds to twice the quark-mixing angle $\alpha$ (green solid line). The other triangles with lighter shade represent the mirror solutions allowed by discrete ambiguities
 , with the corresponding values for $\alpha$ represented by the green dashed lines.}
 \label{fig:isospin_relations}
\end{figure}

The Belle~II detector, complete with its vertex detector, started its colliding beam \mbox{operations} on March 11, 2019.  The sample of electron-positron collisions used in this work corresponds to an integrated luminosity of $62.8\,\si{fb^{-1}}$~\cite{Abudinen:2019osb}  and was collected at the $\Upsilon(4{\rm S})$~\mbox{reso}nance as of July 1, 2020.

This \mbox{document} reports on the first Belle II measurement of branching fraction and  longitudinal polarization fraction of  $\PBplus\to \Prhoplus\Prhozero$ decays,  reconstructed in their $\Pgpp\Pgpz(\to\gamma\gamma)\Pgpp\Pgpm$ final state.
Charge-conjugate processes are implied.\par
All analysis choices and procedures are developed and finalized in simulated and control-sample data before examining experimental data in the signal region.\par Pions in the final states and the large width of $\rho$ mesons offer reduced distinctive features against dominant backgrounds from $e^+e^- \to q\bar{q}$ events, where $q$ indicates any quark of the first or second family (continuum). Isolating a significant, low-background signal is therefore the main challenge of the analysis. We achieve this through an optimization of the selection requirements followed by a sample-composition fit, both based on the following observables:
\begin{itemize}
    \item the beam-energy-constrained mass $M_{\rm bc} \equiv \sqrt{(\sqrt{s}/2c^{2})^{2}- (p^{*}_B/c)^2}$, which is the invariant mass of the $B$ candidate where the energy is replaced by the (more precisely known) half of the center-of-mass collision energy, discriminates fully reconstructed $B$ decays from continuum.
    \item the energy difference $\Delta E \equiv E^{*}_{B} - \sqrt{s}/2$ between the total energy of the reconstructed $B$ candidate and half of the collision energy, both in the $\Upsilon$(4S) frame, also discriminates between signal and misreconstructed $B$ decays;
    \item dipion masses for $\Prhoplus\,(\to\pi^+\pi^0)$ and $\Prhozero\,(\to\pi^+\pi^-)$ candidates, offer further discrimination against continuum and misreconstructed $B$ decays;
    \item the cosines of the helicity angles, defined as the angles between the momentum of the positive-charge pion and the direction opposite to the $B^+$ momentum as measured in the $\rho$ rest frame, provide information on the orbital angular momentum of the final state, which can take values of 0, 1, or 2  due to the decay of a spin-0 particle into two spin-1 particles. 
\end{itemize}

\section{The Belle~II detector}
Belle~II is a nearly $4\pi$ particle-physics spectrometer~\cite{Kou:2018nap, Abe:2010sj}, designed to reconstruct the products of electron-positron collisions produced by the SuperKEKB asymmetric-energy collider~\cite{Akai:2018mbz}, located at the KEK laboratory in Tsukuba, Japan. Belle~II comprises several subdetectors arranged around the interaction space-point in a cylindrical geometry. The innermost subdetector is the vertex detector, which uses position-sensitive silicon layers to sample the trajectories of charged particles (tracks) in the vicinity of the interaction region to extrapolate the decay positions of their long-lived parent particles. The vertex detector includes two inner layers of silicon pixel sensors and four outer layers of silicon microstrip sensors. The second pixel layer is currently incomplete and covers only one sixth of azimuthal angle. Charged-particle momenta and charges are measured by a large-radius, helium-ethane, small-cell central drift chamber, which also offers charged-particle-identification information through a measurement of particles' energy-loss by specific ionization. A  time-of-propagation Cherenkov detector surrounding the chamber provides charged-particle identification (PID) in the central detector volume, supplemented by proximity-focusing, aerogel, ring-imaging Cherenkov detectors in the forward regions. A CsI(Tl)-crystal electromagnetic calorimeter allows for energy measurements of electrons and photons.  A solenoid surrounding the calorimeter generates a uniform axial 1.5\,T magnetic field filling its inner volume. Layers of plastic scintillators and resistive-plate chambers, interspersed between the
magnetic flux-return iron plates, allow for identification of $K^0_{\rm L}$ and muons.
The subdetectors most relevant for this work are the silicon vertex detector, drift chamber, the PID detectors, and the electromagnetic calorimeter.

\section{Selection and reconstruction}
\label{sec:selection}

We reconstruct the two-body decay $\PBplus\to \Prhoplus\,(\to\pi^+\pi^0)\,\Prhozero\,(\to\pi^+\pi^-)\,$ and the control channels
$B^+ \to \overline{D}^0 (\to K^+ \pi^-)\, \rho^+\,(\to\pi^+\pi^0)$, for analysis validation; $B^+ \to \overline{D}^0 (\to K^+ \pi^- \pi^0)\, \pi^+$,  for validation of continuum-suppression inputs and optimization of the $\pi^0$ selection; and 
$B^+ \to \overline{D}^0 (\to K^+ \pi^- )\, \pi^+$, to determine the systematic uncertainty associated with the continuum-suppression and PID.

\subsection{Simulated and experimental data}
We use generic simulated data to optimize the event selection and construct the sample-composition fit model for backgrounds. We use signal-only simulated data to model relevant signal features for the sample-composition fit and determine selection efficiencies. 
The generic simulation consists of Monte Carlo (MC) samples that include $e^+e^-\to B^0\overline{B}^0$, $B^+B^-$, $u\bar{u}$, $d\bar{d}$, $c\bar{c}$, and $s\bar{s}$ processes in realistic proportions and corresponding in size to 10  times the $\Upsilon$(4S)~data. In addition, $6\times 10^4$ signal-only $B^+\to\rho^+\rho^0$ events are generated for each polarization~\cite{Lange:2001}. Finally, simplified simulated experiments constructed by randomly sampling the likelihood of the sample-composition fit allow for studying the estimator properties and assessing fit-related systematic uncertainties (discussed later). \par 
We use all 2019--2020  $\Upsilon$(4S) good-quality experimental data collected until July~1, 2020, which correspond to an integrated luminosity of $62.8\,\si{fb^{-1}}$. All events are required to satisfy loose data-skim selection criteria, based on total energy and charged-particle multiplicity in the event, targeted at reducing sample sizes to a manageable level with negligible impact on signal efficiency. All data are processed using the Belle~II analysis software~\cite{Kuhr:2018lps}.

\subsection{Reconstruction and baseline selection}
We form final-state particle candidates by applying loose baseline selection criteria and then combine candidates in kinematic fits consistent with the topology of the decay to reconstruct intermediate states and finally $B$ candidates. \par We reconstruct charged-pion candidates using the most inclusive charged-particle selection restricted to the full polar-angle acceptance in the central drift chamber ($\SI{17}{\degree}<\theta<\SI{150}{\degree}$) and to loose requirements on impact parameters (displacement from the nominal interaction space-point, $|dr|<\SI{0.5}{cm}$ radial and $|dz|<\SI{3.0}{cm}$ longitudinal with respect to the beams) to reduce beam-background-induced tracks, which do not originate from the interaction region.
We reconstruct neutral-pion candidates by combining pairs of photons with energies greater than about $20$\,MeV restricted in diphoton mass to within approximately three times the resolution from the known $\pi^0$ mass and excluding extreme helicity-angle values  to suppress combinatorial background from collinear soft photons. The mass of the $\pi^0$ candidates is constrained to its known value in subsequent kinematic fits.
The resulting $\pi^\pm$ and $\pi^0$ candidates are combined to form $\Prhoplus$ and $\Prhozero$ candidates by requiring $0.52 < m(\pi^+\pi^{0,-}) < 1.06$\,GeV/$c^2$. We apply a selection on the helicity angle of the $\Prhoplus$ candidates, to suppress candidates formed with soft neutral pions, which usually originate from background.  Pairs of $\Prhoplus$ and $\Prhozero$~candidates are then combined to form $B$-meson candidates. The resulting $B$ candidates are combined through kinematic fits of the entire decay chain with appropriate constraints. In addition, we reconstruct the vertex of the accompanying tag-side $B$ mesons using  all tracks in the tag-side and identify the flavor, which is used as  input to the continuum-background discriminator, using the Belle II flavor tagger~\cite{Abudinen:2018}.
The reconstruction of the control channels is conceptually similar and includes additional charmed-meson invariant mass restrictions when appropriate.
\par
Simulation is used to identify and suppress contamination from peaking backgrounds, that is, misreconstructed events clustering in the signal region $M_{\rm bc} > 5.27$\,GeV/$c^2$ and \mbox{$-0.15 < \Delta E < 0.15$ GeV}.  
Peaking backgrounds from decays with intermediate charmed\break 
resonances are suppressed by requiring 
$\vert m(\Pgppm\Pgpmp) - m_{\PDzero}\vert > \SI{18}{MeV}/c^{2}$ and\break 
$\vert m(\Pgppm\Pgpmp\Pgpz) - m_{\PDzero}\vert > \SI{32}{MeV}/c^{2}$, where $m_{\PDzero}$ is the known \PDzero mass~\cite{Zyla:2020zbs}, and\break 
$\vert m(\Pgppm\Pgpmp) - \SI{1.78}{GeV}/c^2\vert > \SI{18}{MeV}/c^{2}$ and 
$\vert m(\Pgppm\Pgpmp\Pgpz) - \SI{1.78}{GeV}/c^2\vert > \SI{32}{MeV}/c^{2}$, where 1.78 GeV/$c^2$ is the position of the charmed signal when a kaon is misidentified as a pion.


\subsection{Continuum suppression}
The main challenge in observing significant charmless signals is the large contamination from continuum background. We use a binary boosted decision-tree classifier that non-linearly combines 39 variables known to provide statistical discrimination between $B$-meson signals and continuum and to be loosely correlated, or uncorrelated,  with $\Delta E$ and $M_{\rm bc}$. These variables include quantities associated with event topology (global and signal-only angular configurations), flavor-tagger information, vertex separation and uncertainty information, and kinematic-fit quality information. We train the classifier to identify statistically significant signal and background features using unbiased simulated samples. 

\begin{figure}[h!]
 \centering
    \centering
    \subfigure{\includegraphics[width=0.49\textwidth]{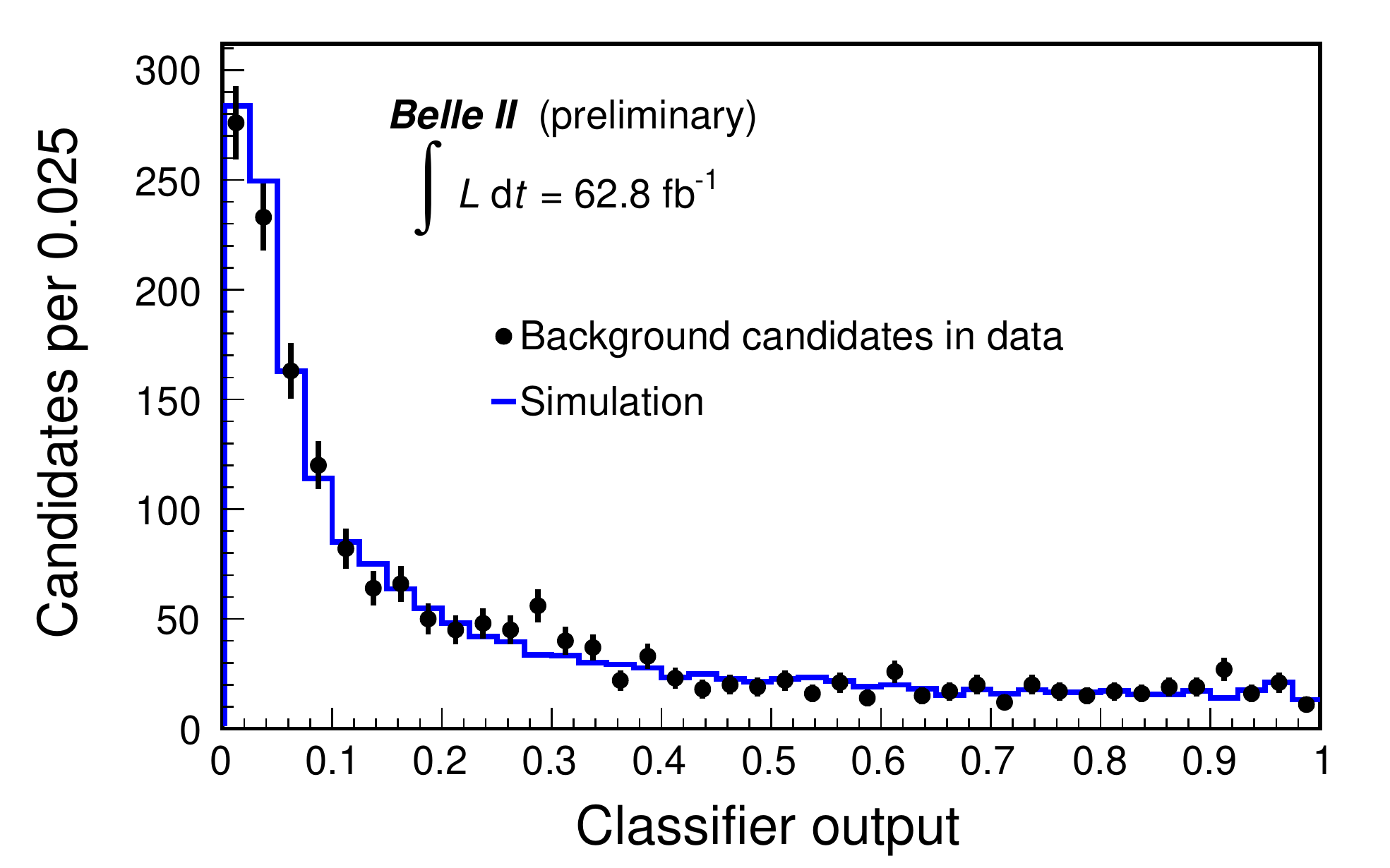}} \hfill
    \subfigure{\includegraphics[width=0.49\textwidth]{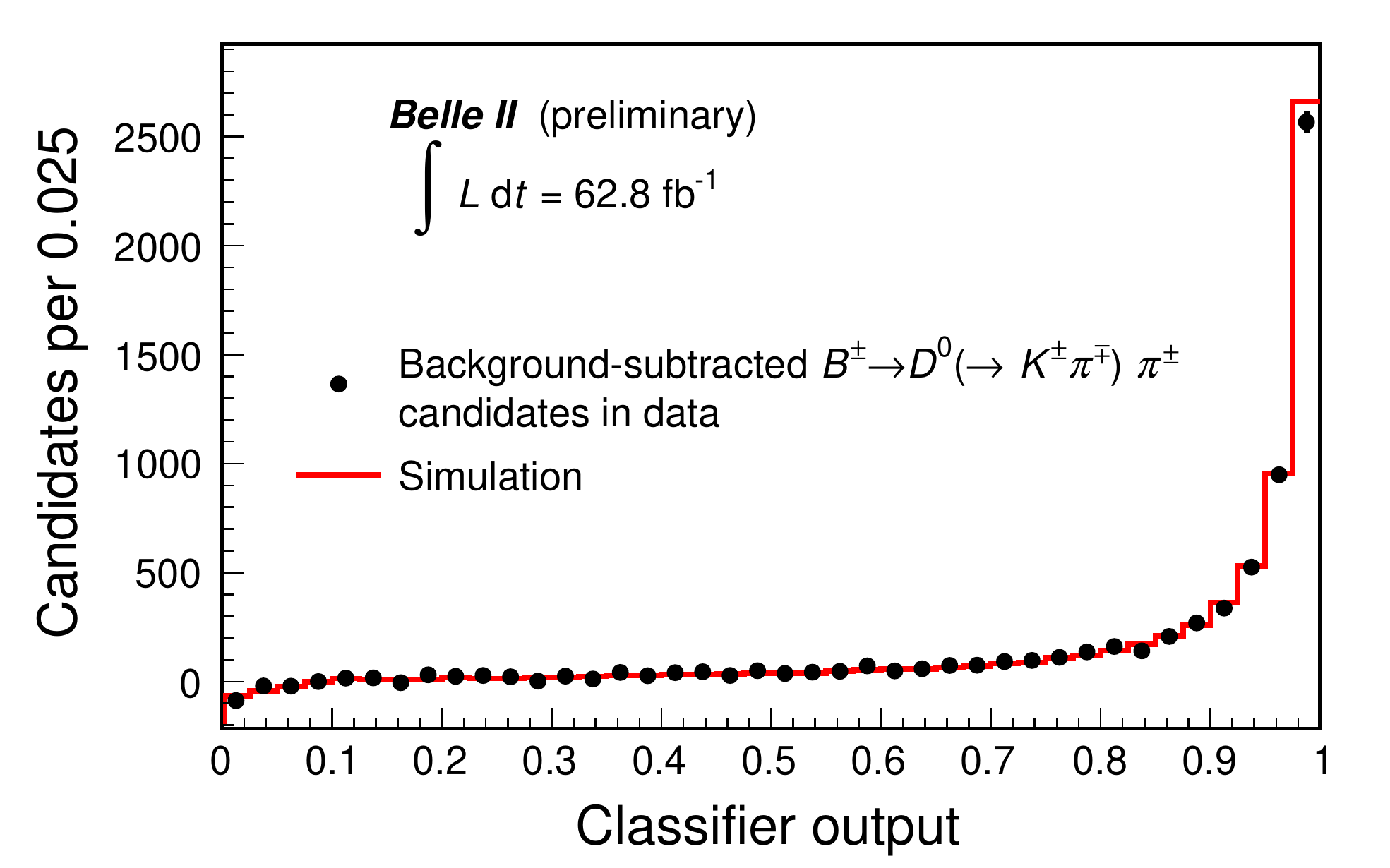}} \\
 \caption{Data-simulation comparison of the output of the boosted decision-tree classifier on (left)~sideband and (right)~sideband-subtracted $\PBplus\to\APD^{0}(\to \PKp\Pgpm)\,\Pgpp$~candidates in the signal region.}
 \label{fig:outputData_Kpi}
\end{figure}

We validate the input and output distributions of the classifier by comparing  data with simulation using control samples.
Figure~\ref{fig:outputData_Kpi} shows the distribution of the output for  \mbox{$\PBplus\to\APD^{0}(\to \PKp\Pgpm)\,\Pgpp$}~candidates. No inconsistency is observed. 


\section{Optimization of the signal selection}
\label{sec:yields}
We optimize the selection using simulated and control-sample data. We vary simultaneously the selection criteria on continuum-suppression output and charged-particle identification information to maximize ${\rm S}/\sqrt{{\rm S}+{\rm B}}$, where ${\rm S}$ and ${\rm B}$ are expected signal and background yields, respectively, as estimated in the signal region of simulated samples. The optimal selection criterion on the continuum-suppression output is $C'_{\rm FBDT} > 0.97$, which removes approximately 98\% of continuum and retains approximately 41\% of signal. The optimal PID criterion has 96\% efficiency for pions with 7\% kaon misidentification rate. In addition, the $\pi^0$ selection is optimized independently by using control \mbox{$B^+ \to \overline{D}^0(\to K^+\pi^-\pi^0)\pi^+$} decays.

\section{Determination of signal yields}
\label{sec:yields}
More than one candidate per event often populates the resulting sample, with average multiplicity of 1.3. We restrict to one candidate per event by selecting the $\pi^0$ candidate with lowest $\chi^2$ value of its mass-constrained diphoton fit, and then selecting the $B$ candidate with lowest vertex-fit $\chi^2$. \par Signal yields are determined with maximum likelihood fits of the unbinned distributions of $\Delta E$, continuum-suppression decision-tree output, $\rho$ candidate dipion masses, and cosines of the helicity angles of the $\rho$ candidates, for candidates in the signal region $M_{\rm bc} > 5.27$\,GeV/$c^2$, $-0.15 < \Delta E < 0.15$ GeV, and $0.52 < m(\pi^+\pi^{-,0}) < 1.06$\,GeV/$c^2$.
The joint, six-dimensional  probability density function is approximated as the product of one-dimensional probabilty density functions.
Sample components used in the fit are transversely-polarized signal, longitudinally-polarized signal, $\PB\overline{\PB}$~background, and continuum, whose yields are determined by the fit. Self cross-feed (i.e., incorrectly reconstructed candidates in signal events), background from $\PB\to f_0(980)X$ decays and nonresonant $\PB\to\Prho\pi\pi$ decays, are determined by the fit within Gaussian constraints based on expectations from previous measurements or phenomenological considerations. Nonresonant $B^+ \to \pi^+\pi^0\pi^+\pi^-$ contributions are assumed to be negligible.\par 
Fit models are determined empirically from simulation. To account for modest differences between data and simulation, we include a global additive shift of the peak positions and a global multiplicative width scale-factor for the $\Delta E$ and $m(\pi^+\pi^0)$ distributions, determined as indicated by likelihood-ratio tests on control channels in data. 
We use a sum of three Gaussian for all the continuum-suppression shapes and for the signal $\Delta E$ shapes, relativistic P-wave Breit-Wigner distributions for the signal $m(\pi^+\pi^{0,-})$ shapes, and a 2nd-order polynomial function for the $\Delta E$ continuum background. All other shapes are modeled with histogram templates from simulation.\par
\begin{figure}[htb]
 \centering
    \centering
    \subfigure{\includegraphics[width=0.425\textwidth]{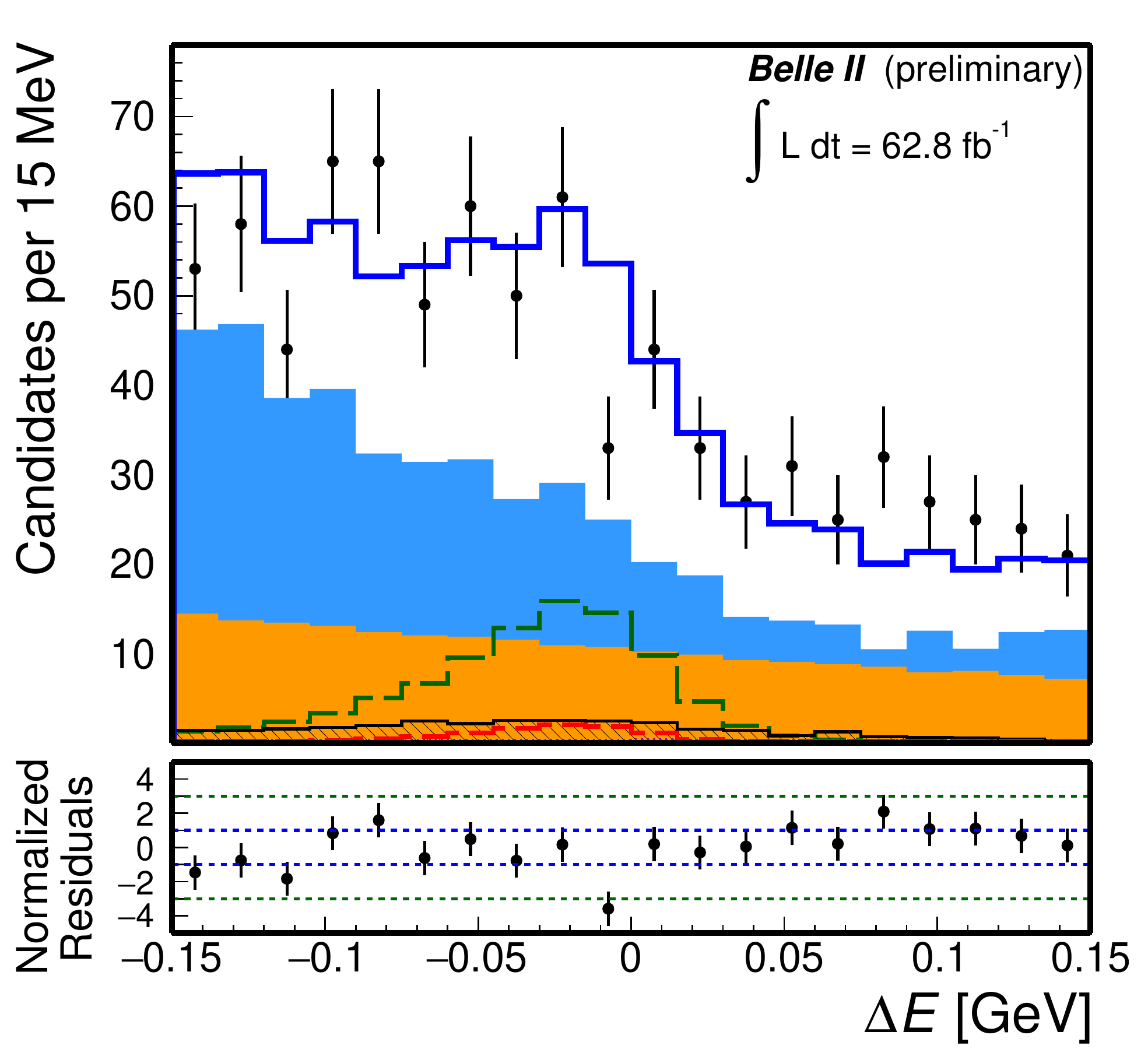}}
    \subfigure{\includegraphics[width=0.425\textwidth]{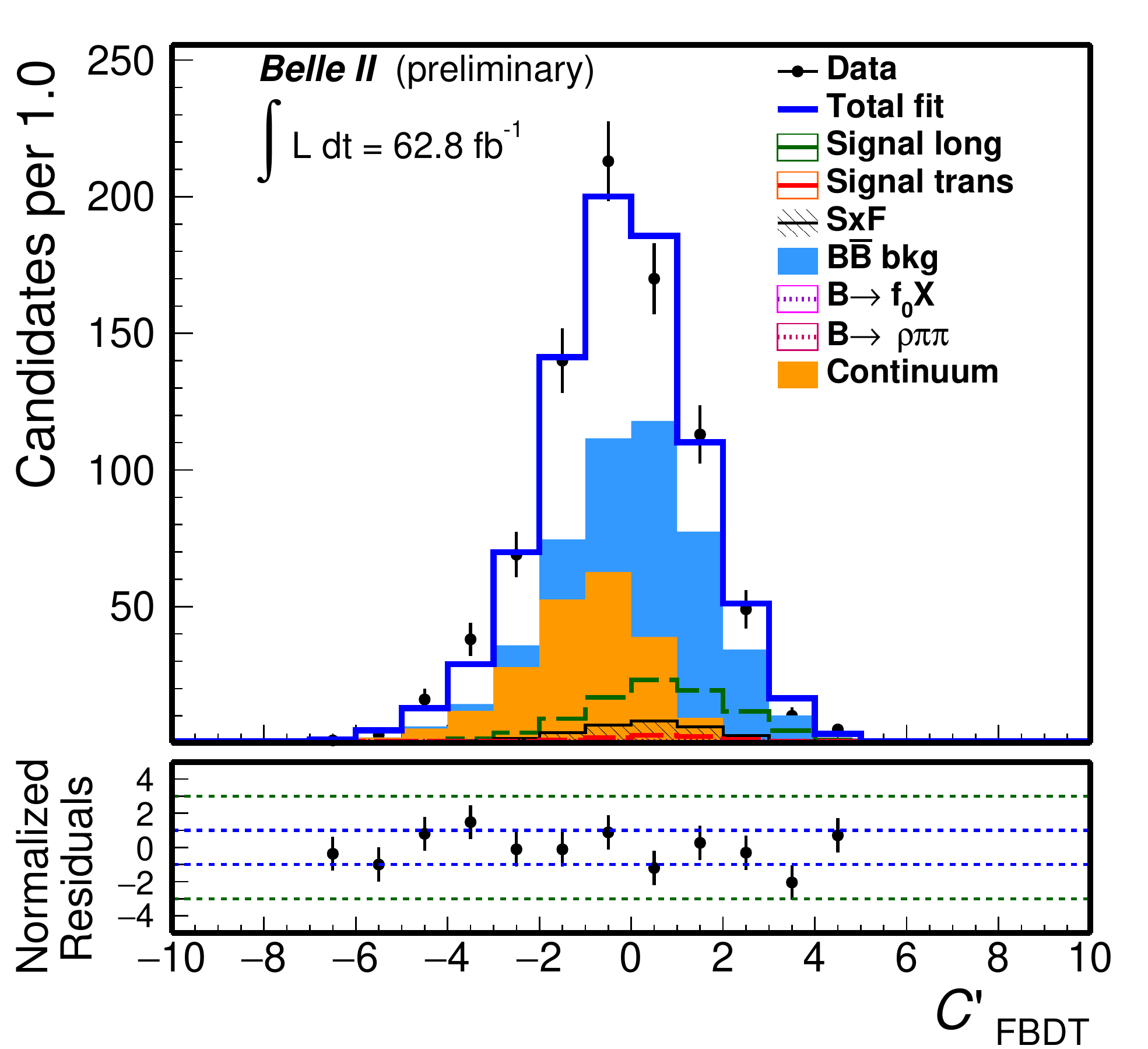}}
    \subfigure{\includegraphics[width=0.425\textwidth]{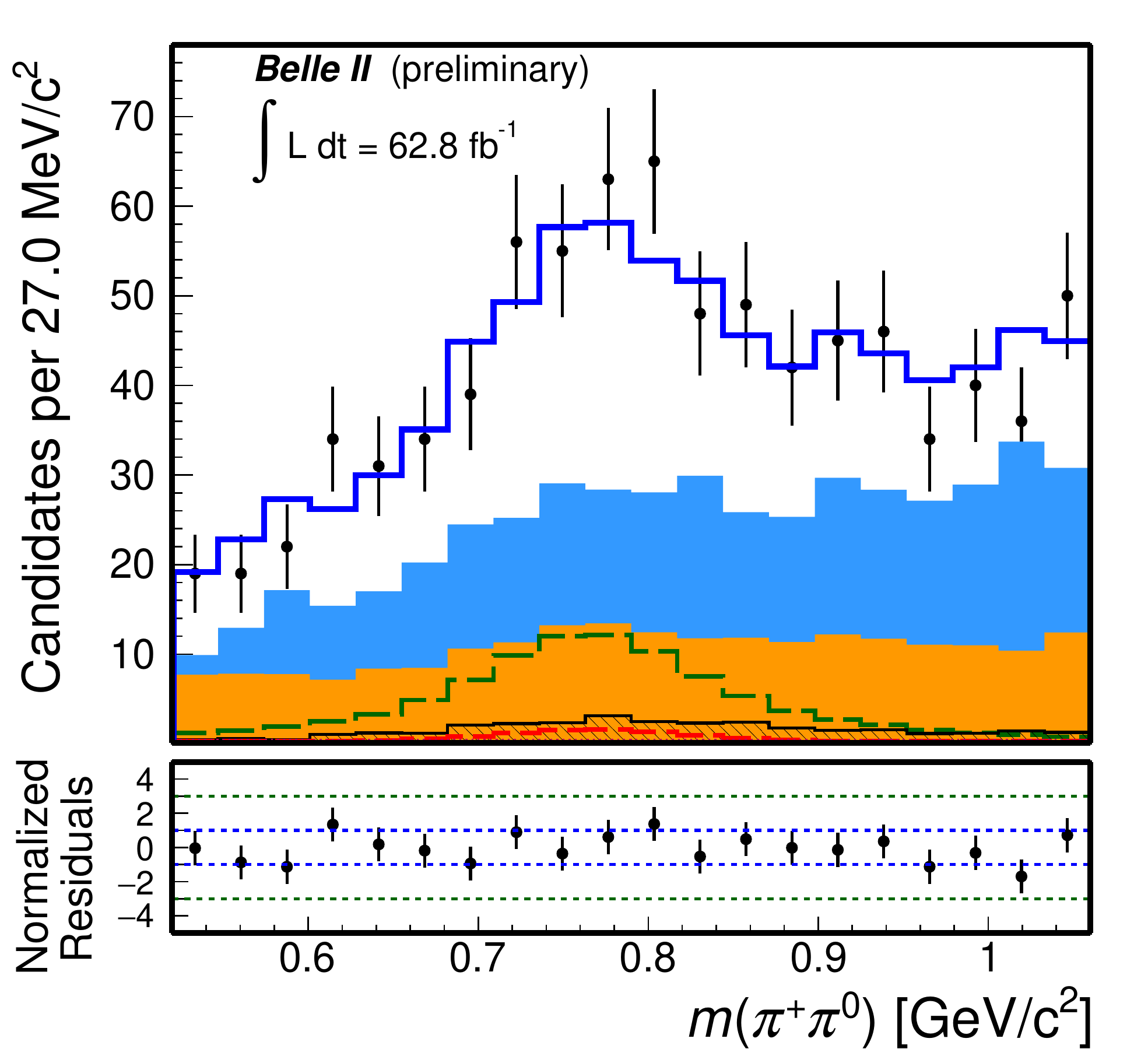}}
    \subfigure{\includegraphics[width=0.425\textwidth]{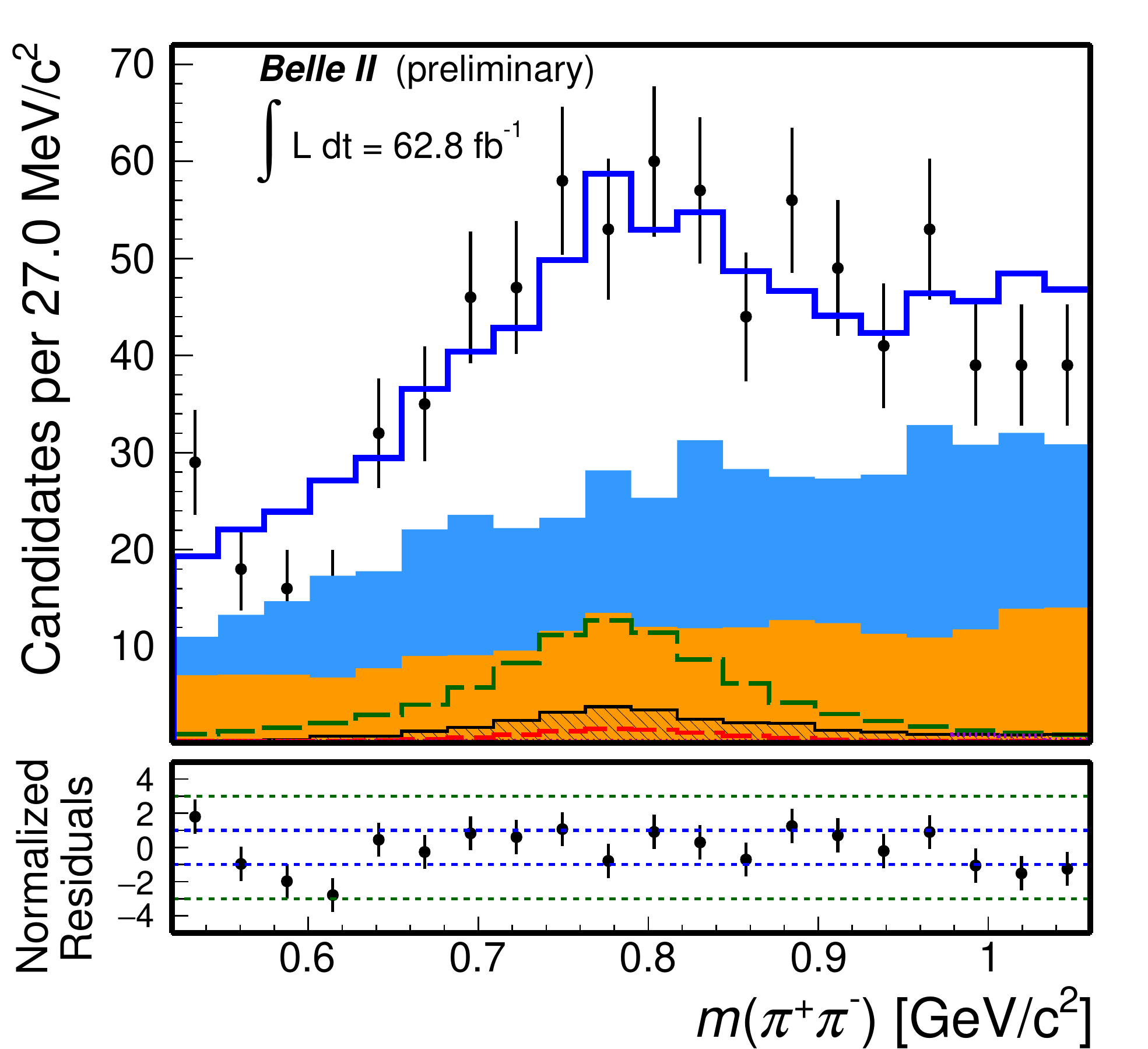}}
    \subfigure{\includegraphics[width=0.425\textwidth]{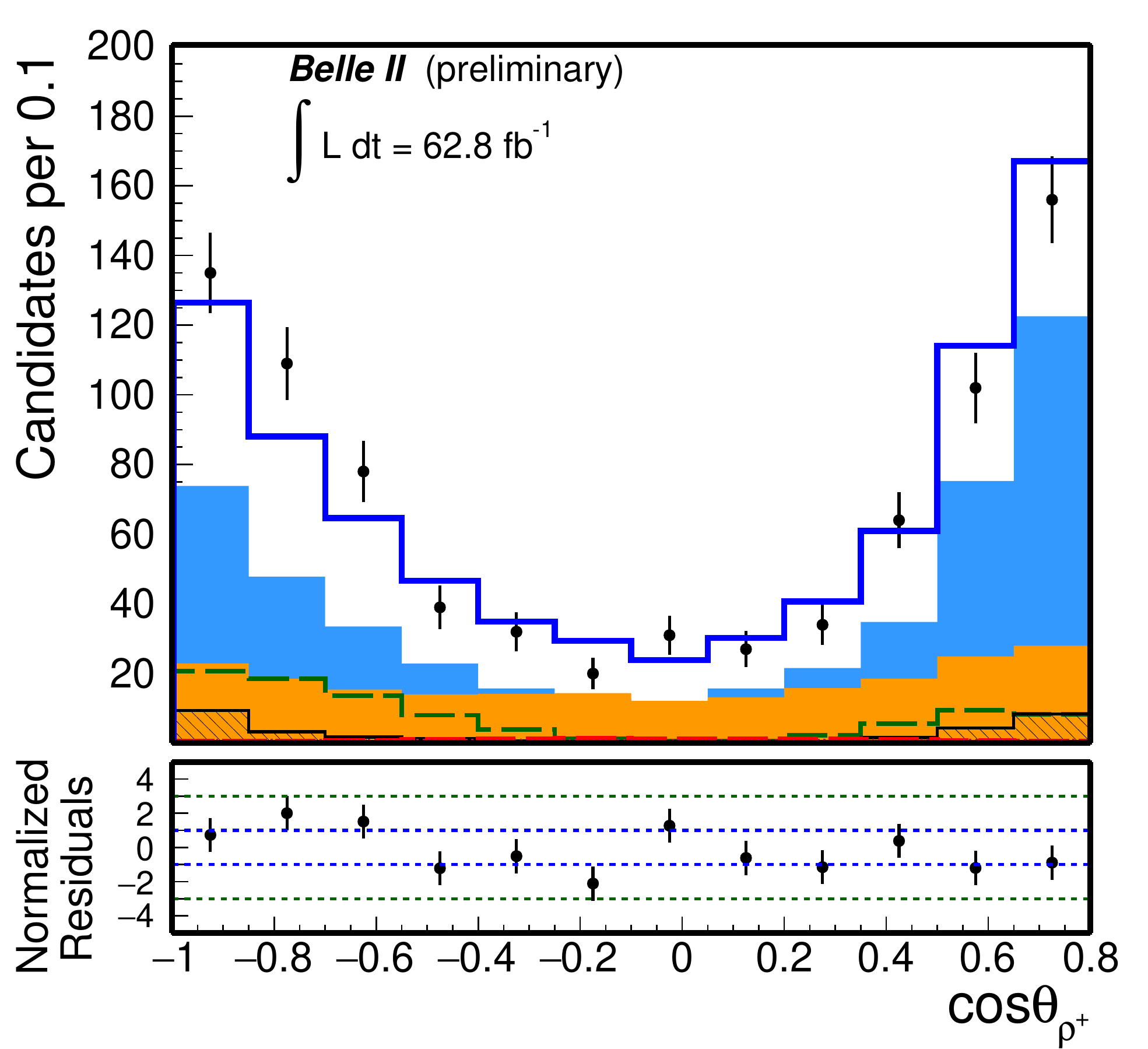}}
    \subfigure{\includegraphics[width=0.425\textwidth]{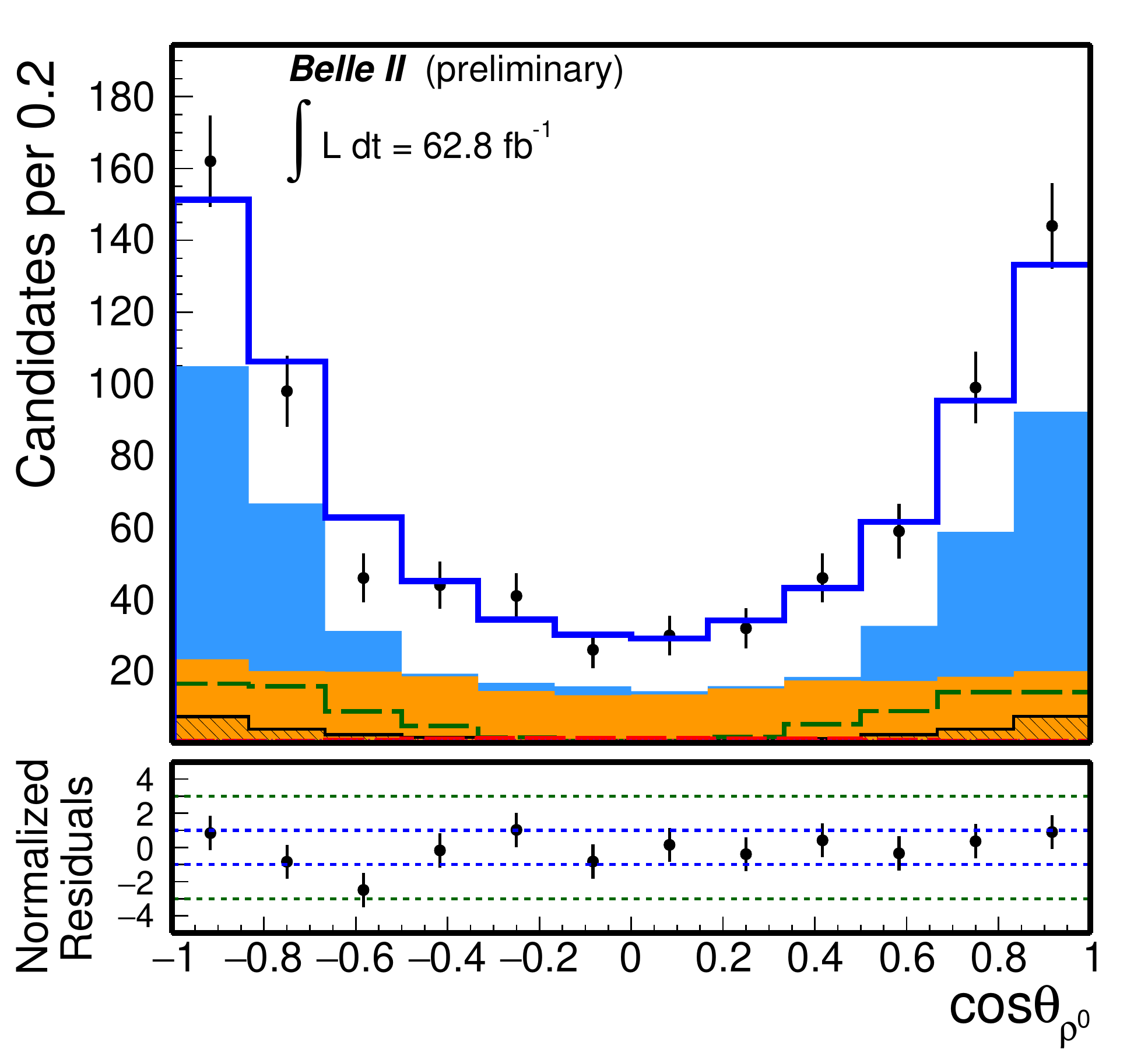}}
 \caption{Distributions of (top left) $\Delta E$, (top right) continuum-suppression decision-tree output, (middle left) $m(\pi^+\pi^0$), (middle right) $m(\pi^+\pi^-$), cosine of the helicity angle of (bottom left) $\rho^+$ and (bottom right) $\rho^0$ for $B^+ \to \rho^+\rho^0$  candidates reconstructed in 2019--2020 Belle~II data selected through the baseline criteria with an optimized continuum-suppression and pion-enriching selection, and further restricted to $M_{\rm bc} > 5.27$\,GeV/$c^2$. Vetoes for peaking backgrounds are applied. Fit projections are overlaid.}
 \label{fig:fits}
\end{figure}

Distributions of fit observables in Belle II data are shown in Figure~\ref{fig:fits} with projections overlaid. 
Broadly-peaking signal structures are visible in the energy-difference and dipion mass distributions, otherwise dominated by the smooth continuum and $B\overline{B}$ background components.
The results of the fit are summarized in Table~\ref{tab:FitSummary}.

\begin{table}[!ht]
    \centering
\caption{Fit results in data.} 
\begin{tabular}{l  c}
\hline\hline
Component & \multicolumn{1}{c}{Fit result} \\\hline
 Longitudinal signal $N_L$     &  $93\pm16$ \\ 
 Transverse signal $N_T$	   &   $ 11^{+9}_{-7} $\\
 Self cross-feed fraction	   &  $ 0.31\pm0.03 $\\
 $B\overline{B}$ background	   &  $ 488\pm37 $\\
 $B^+ \to f_0 \rho^+$ decays &   $ 2\pm3 $	\\
 $B^+ \to \rho^{(0,+)}\pi^{(0,-)}\pi^+$ decays &   $ -11\pm5 $\\
 Continuum background	   &  $ 212\pm30 $ \\
\hline
\end{tabular}
    \label{tab:FitSummary}
\end{table}{}

\clearpage

\section{Efficiencies and corrections}

\label{sec:efficiencies}

The raw event yields observed in data are divided by selection and reconstruction efficiencies to obtain the branching fraction and  the fraction of longitudinally polarized events. The efficiencies, $6.7\%$ and $12.0\%$ for longitudinal and transverse signal, respectively, are determined from simulation with statistical uncertainties of $0.1\%$. We perform consistency checks on control samples of data and assess systematic uncertainties for those factors contributing to the above efficiencies where simulation may not accurately describe data.

\section{Determination of branching fraction and fraction of longitudinally polarized decays}

We determine the branching fraction and the fraction of longitudinally polarized decays as 

\begin{equation*}
    \mathcal{B} = \frac{\frac{N_{L}}{\varepsilon_{L}} + \frac{N_{T}}{\varepsilon_{T}}}{2 N_{\PB\APB}},\ \ f_L = \frac{\frac{N_{L}}{\varepsilon_{L}}}{\frac{N_{L}}{\varepsilon_{L}} + \frac{N_{T}}{\varepsilon_{T}}}
\end{equation*}
where $N_L$ and $N_T$ are the signal yields determined by the fit, $\varepsilon_L$ and $\varepsilon_T$ are the corresponding selection and reconstruction efficiencies, and $N_{\PB\APB}$ is the number of produced $\PBplus\PBminus$ pairs, corresponding to $35.8\times 10^6$. We obtain the number of ${\PB\APB}$~pairs from the measured integrated luminosity and the \mbox{$\Pep\Pem\to\Upsilon(4{\rm S})$}~cross section~$(1.110 \pm 0.008)\,$nb~\cite{Bevan:2014iga}, assuming that the $\Upsilon(4{\rm S})$ decays exclusively to ${\PB\APB}$~pairs, and  \mbox{$f^{00} = 0.487\pm 0.010\pm 0.008$} for the \mbox{$\Upsilon(4{\rm S})\to\PBzero\APBzero$}~branching fraction~\cite{Aubert:2005bq}. Since our results (Table I) are consistent with full longitudinal polarization of the vector mesons, we assume 100\% for the  \mbox{$\Prhoplus \to \pi^+\pi^0$} and $\Prhozero \to \pi^+\pi^-$ branching fractions.

\section{Systematic uncertainties}

We consider several systematic effects, assumed to be independent, and add in quadrature the corresponding uncertainties. An overview of the effects considered follows. A summary of the fractional size of systematic uncertainties is listed in Table~\ref{tab:Systematics_overview}.

\subsection{Tracking efficiency} 
We assess a systematic uncertainty associated with possible data-simulation discrepancies in the reconstruction of charged particles~\cite{Bertacchi:2020eez}.
The tracking efficiency in data agrees with the value observed in simulation within a $0.69\%$ uncertainty, which we (linearly) add as a systematic uncertainty for each final-state charged particle.

\subsection{\Pgpz~reconstruction efficiency} 
We assess a systematic uncertainty associated with possible data-simulation discrepancies in the $\Pgpz$ reconstruction and selection using the decays \mbox{$B^0 \to D^{*-}(\to \overline{D}^0 (\to K^+ \pi^- \pi^0)\, \pi^-)\, \pi^+$} and \mbox{$B^0 \to D^{*-}(\to \overline{D}^0 (\to K^+ \pi^-)\, \pi^-)\, \pi^+$} where the selection of charged particle is identical and all distributions are weighted so as the $\pi^0$ momentum matches the $\pi^0$ momentum in charmless channels. We compare the yields obtained from fits to the $\Delta E$ distribution of reconstructed $\PB$~candidates 
and obtain an efficiency in data that agrees with the value observed in simulation within an uncertainty, which is used as a systematic uncertainty. This is the second largest contributor to the systematic uncertainties.

\subsection{Particle-identification and continuum-suppression efficiencies} We identify possible data-simulation discrepancies in particle identification and continuum-suppression distributions using the control channel 
\mbox{$\PBplus\to\APDzero(\to\PKp\Pgpm)\,\Pgpp$}.
We find that the selection efficiencies obtained in data and simulation agree within their statistical uncertainty, which is taken as a systematic uncertainty.

\subsection{Number of $\PB\APB$~pairs} We  assign  a  systematic  uncertainty associated with the uncertainty on the number of $\PB\APB$~pairs,  which  includes  the uncertainty  on  cross-section,  integrated  luminosity~\cite{Abudinen:2019osb}, and  potential  shifts  from  the peak center-of-mass energy during the run periods.

\subsection{Signal and background modeling} We quote, as the systematic uncertainty for possible signal or background mismodelings, the difference between the average of results of the sample-composition fit performed on ensembles of  simplified simulated experiments generated with the baseline model and various alternate models. Uncertainties on branching fraction and longitudinal polarization fraction range from 0.1\% to 0.5\%.

\subsection{Single candidate selection}

We assess a systematic uncertainty on the single-candidate selection by re-doing the analysis selecting randomly one $\PBplus$ candidate per event. The difference between results is taken as a systematic uncertainty.

\subsection{Fit biases}

We include in the systematic uncertainty statistical biases observed in fits of ensembles of simplified simulated experiments. 

\subsection{Simulation mismodeling}

The largest systematic uncertainty is due to significant data-simulation discrepancies observed in distributions of fit observables for candidates populating 
signal sidebands, which could not be conclusively attributed to shape/acceptance mismodelings or poorly simulated sample composition. A systematic uncertainty based on the deviation of results in fits to simulated ensembles that mirror the observed discrepancies generously covers any possible effect. 

\begin{table}[h]
\centering
\footnotesize
\caption{Summary of the (fractional) systematic uncertainties of the branching-fraction and longitudinal-polarization-fraction measurements.}
\begin{tabular}{l c c}
\hline\hline
 Source & $\mathcal{B}$ & $f_L$\\
\hline
Tracking             & 2.07 \% & - -\\
$\Pgpz$ efficiency      & 13.0 \%  & - -\\
PID and continuum-supp. eff. & 0.8 \% & - - \\
One-candidate selction & 0.85 \% & 0.87 \% \\
$N_{B\bar{B}}$      &  1.4 \%  &  - - \\
Signal model & 0.5 \%  & 0.1 \%  \\
Continuum bkg. model &  0.5 \%  & 0.1 \% \\
$\PB\APB$ bkg model &  0.5 \%  & 0.1 \%  \\
Fit biases & 1.5 \% & 1.5 \%  \\
Data-MC mismodeling & 14.2 \% & 1.3 \% \\
\hline
Total  & 19.5 \%  & 2.2 \%  \\
\hline\hline
\end{tabular} 

\label{tab:Systematics_overview}
\end{table}

\section{Results and summary}
\label{sec:summary}
We report on a measurement of the branching fraction  and longitudinal polarization fraction of $B^+\to \rho^+\rho^0$ decays. We use a sample of 2019 and 2020 data collected by the Belle II experiment and corresponding to $62.8$\,fb$^{-1}$ of integrated luminosity. We use simulation to determine optimized event selections
and fit the $\Delta E$, continuum-background suppression, invariant masses and angular distributions of the resulting samples to determine a $B^+\to \rho^+\rho^0$ signal yield of $104\pm16$ decays. The signal yields are corrected for efficiencies determined from simulation and control data samples to obtain 
\begin{center}
$\mathcal{B}(B^+ \to \rho^+\rho^0) = [20.6 \pm 3.2(\rm stat) \pm 4.0(\rm syst)]\times 10^{-6}$, \mbox{and}  \\
\end{center}
\begin{center}
$f_L(B^+ \to \rho^+\rho^0) = 0.936  ^{+0.049}_{-0.041} (\rm stat)\pm 0.021(\rm syst)$.
\end{center}

These are the first measurements of  $\PBplus\to\Prhoplus\Prhozero$ decay properties reported by Belle~II. 
Results are compatible with previous determinations~\cite{Zhang:2003up,BaBar:2009rmk} and show performance superior to early Belle results~\cite{Zhang:2003up}.

\clearpage

\section*{Acknowledgments}

We thank the SuperKEKB group for the excellent operation of the
accelerator; the KEK cryogenics group for the efficient
operation of the solenoid; the KEK computer group for
on-site computing support; and the raw-data centers at
BNL, DESY, GridKa, IN2P3, and INFN for off-site computing support.
This work was supported by the following funding sources:
Science Committee of the Republic of Armenia Grant No. 20TTCG-1C010;
Australian Research Council and research grant Nos.
DP180102629, 
DP170102389, 
DP170102204, 
DP150103061, 
FT130100303, 
FT130100018,
and
FT120100745;
Austrian Federal Ministry of Education, Science and Research,
Austrian Science Fund No. P 31361-N36, and
Horizon 2020 ERC Starting Grant no. 947006 ``InterLeptons''; 
Natural Sciences and Engineering Research Council of Canada, Compute Canada and CANARIE;
Chinese Academy of Sciences and research grant No. QYZDJ-SSW-SLH011,
National Natural Science Foundation of China and research grant Nos.
11521505,
11575017,
11675166,
11761141009,
11705209,
and
11975076,
LiaoNing Revitalization Talents Program under contract No. XLYC1807135,
Shanghai Municipal Science and Technology Committee under contract No. 19ZR1403000,
Shanghai Pujiang Program under Grant No. 18PJ1401000,
and the CAS Center for Excellence in Particle Physics (CCEPP);
the Ministry of Education, Youth and Sports of the Czech Republic under Contract No.~LTT17020 and 
Charles University grants SVV 260448 and GAUK 404316;
European Research Council, 7th Framework PIEF-GA-2013-622527, 
Horizon 2020 ERC-Advanced Grants No. 267104 and 884719,
Horizon 2020 ERC-Consolidator Grant No. 819127,
Horizon 2020 Marie Sklodowska-Curie grant agreement No. 700525 `NIOBE,' 
and
Horizon 2020 Marie Sklodowska-Curie RISE project JENNIFER2 grant agreement No. 822070 (European grants);
L'Institut National de Physique Nucl\'{e}aire et de Physique des Particules (IN2P3) du CNRS (France);
BMBF, DFG, HGF, MPG, and AvH Foundation (Germany);
Department of Atomic Energy under Project Identification No. RTI 4002 and Department of Science and Technology (India);
Israel Science Foundation grant No. 2476/17,
United States-Israel Binational Science Foundation grant No. 2016113, and
Israel Ministry of Science grant No. 3-16543;
Istituto Nazionale di Fisica Nucleare and the research grants BELLE2;
Japan Society for the Promotion of Science,  Grant-in-Aid for Scientific Research grant Nos.
16H03968, 
16H03993, 
16H06492,
16K05323, 
17H01133, 
17H05405, 
18K03621, 
18H03710, 
18H05226,
19H00682, 
26220706,
and
26400255,
the National Institute of Informatics, and Science Information NETwork 5 (SINET5), 
and
the Ministry of Education, Culture, Sports, Science, and Technology (MEXT) of Japan;  
National Research Foundation (NRF) of Korea Grant Nos.
2016R1\-D1A1B\-01010135,
2016R1\-D1A1B\-02012900,
2018R1\-A2B\-3003643,
2018R1\-A6A1A\-06024970,
2018R1\-D1A1B\-07047294,
2019K1\-A3A7A\-09033840,
and
2019R1\-I1A3A\-01058933,
Radiation Science Research Institute,
Foreign Large-size Research Facility Application Supporting project,
the Global Science Experimental Data Hub Center of the Korea Institute of Science and Technology Information
and
KREONET/GLORIAD;
Universiti Malaya RU grant, Akademi Sains Malaysia and Ministry of Education Malaysia;
Frontiers of Science Program contracts
FOINS-296,
CB-221329,
CB-236394,
CB-254409,
and
CB-180023, and SEP-CINVESTAV research grant 237 (Mexico);
the Polish Ministry of Science and Higher Education and the National Science Center;
the Ministry of Science and Higher Education of the Russian Federation,
Agreement 14.W03.31.0026;
University of Tabuk research grants
S-0256-1438 and S-0280-1439 (Saudi Arabia);
Slovenian Research Agency and research grant Nos.
J1-9124
and
P1-0135; 
Agencia Estatal de Investigacion, Spain grant Nos.
FPA2014-55613-P
and
FPA2017-84445-P,
and
CIDEGENT/2018/020 of Generalitat Valenciana;
Ministry of Science and Technology and research grant Nos.
MOST106-2112-M-002-005-MY3
and
MOST107-2119-M-002-035-MY3, 
and the Ministry of Education (Taiwan);
Thailand Center of Excellence in Physics;
TUBITAK ULAKBIM (Turkey);
Ministry of Education and Science of Ukraine;
the US National Science Foundation and research grant Nos.
PHY-1807007 
and
PHY-1913789, 
and the US Department of Energy and research grant Nos.
DE-AC06-76RLO1830, 
DE-SC0007983, 
DE-SC0009824, 
DE-SC0009973, 
DE-SC0010073, 
DE-SC0010118, 
DE-SC0010504, 
DE-SC0011784, 
DE-SC0012704, 
DE-SC0021274; 
and
the Vietnam Academy of Science and Technology (VAST) under grant DL0000.05/21-23.

\bibliography{belle2}
\bibliographystyle{belle2-note}

\end{document}